\documentclass[aps,prl,twocolumn,groupedaddress]{revtex4}

\usepackage{graphicx}

\newcommand{\rs}{\rm \scriptscriptstyle}

\begin{document}

\title{Supersolid versus Phase Separation in Atomic Bose-Fermi Mixtures}

\author{H.P.\ B\"uchler}
\author{G.\ Blatter}
\affiliation{Theoretische Physik, ETH-H\"onggerberg, CH-8093
Z\"urich, Switzerland}

\date{\today}

\begin{abstract}
  We show that a two-dimensional atomic mixture of Bosons and
  Fermions cooled into their quantum degenerate states and subject to
  an optical lattice develops a supersolid phase characterized
  by the simultaneous presence of a non-trivial crystalline order
  and phase order. This transition is in competition with a phase
  separated ground state. We determine the phase diagram of
  the system and propose an experiment allowing for the
  observation of the supersolid phase.
\end{abstract}


\maketitle

Cooling atoms to the nK regime allows for the realization and
study of new thermodynamic phase transitions and their associated
phases, with an interesting synergy emerging between the fields of
quantum atom optics and condensed matter physics. Recent trends
are the study of the superfluid to Mott-insulator phase transition
appearing in cold bosonic systems subject to an optical lattice
\cite{jaksch98,greiner02} and the striving for the realization of
a BCS-type condensate in a fermionic system
\cite{houbiers97,heiselberg00}. In this letter, we investigate the
possibility to realize a non-trivial supersolid phase in a mixed
boson-fermion system sympathetically cooled into their
corresponding quantum degenerate states \cite{truscott01,roati02}.
We identify a promising system where this novel phase can be
observed and determine the relevant phase diagram.

Supersolids simultaneously exhibit two types of order which
usually appear in competition to each other --- these are the
diagonal long-range order (DLRO) associated with the periodic
density modulation in a crystal and the off-diagonal long-range
order (ODLRO) associated with the phase order in the condensate
\cite{penrose56}. Supersolids have been proposed to exist in the
strongly interacting $^4$He system \cite{andreev69,leggett70},
where experimental results are still hotly debated
\cite{lengua90,meisel92}, and in various model systems describing
interacting Bosons on a lattice and analyzed numerically
\cite{batrouni95,frey97,goral02}. Here, we investigate the
possibility to use a specifically tuned boson-fermion mixture to
realize a supersolid phase in a controlled experiment.

The basic idea underlying our scheme is to share tasks between the
fermions and the bosons: the fermions are tuned through a density
wave instability establishing crystalline order (DLRO), while the
condensate bosons provide the off-diagonal long range order
(ODLRO). The interaction with the fermions imprints an additional
density modulation also in the bosonic density field, hence
resulting in a supersolid phase. In order to trigger a density
wave instability in the fermions, we confine the mixed
boson-fermion system to two dimensions and subject it to an
optical lattice providing perfect Fermi surface nesting at
half-filling \cite{gruner}. Note, that the resulting crystalline
order relevant for the DLRO component in the supersolid is not due
to the density modulation enforced by the optical lattice but is
the superstructure triggered by the density wave instability.

\begin{figure}[hbtp]
\vspace{-0.cm}
\includegraphics[scale=0.58]{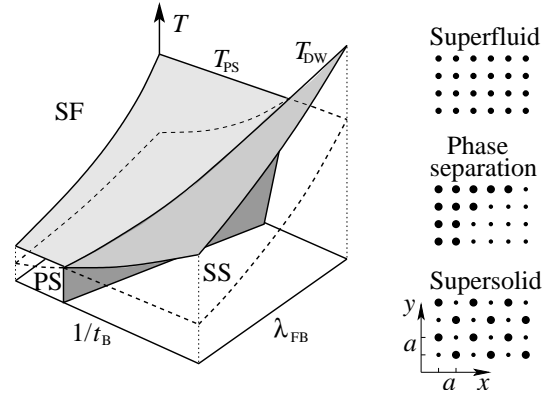}
\vspace{-0.2cm}
  \caption{ \label{phasedia} Left: Sketch of the
  $\lambda_{\rs FB}$-$1/t_{\rs B}$-$T$ phase diagram.
  For $T_{\rs DW}>T_{\rs PS}$ the low temperature phase
  at $T<T_{\rs DW}$ is a supersolid (SS), while phase
  separation (PS) emerges at $T<T_{\rs PS}$ for $T_{\rs PS}>T_{\rs DW}$.
  Right: bosonic density $n_{\rs B}(x,y)$ for the superfluid
  (SF), phase separated, and supersolid phases.}
\end{figure}

The supersolid transition triggered by the fermions competes with
an instability towards phase separation in the boson system
\cite{molmer98,amoruso98}. Given the dimensionality and the
lattice geometry of our system, the presence of van Hove
singularities strongly enhances the tendency towards phase
separation and produces new and interesting features in this
transition: an arbitrary weak interaction between the bosons and
the fermions is sufficient to drive the phase separation at low
temperatures. However, proper tuning of the parameters allows to
supersede this phase separation through the supersolid transition
shown in Fig.~\ref{phasedia}. In the following, we first
investigate the two instabilities towards phase separation and
density wave formation and then analyze the supersolid phase
within a mean-field scheme. We focus on the weak coupling limit
between the bosons and the fermions, which excludes a demixing in
a repulsive fermion-boson system along the lines discussed in
Ref.\ \cite{viverit00}.

\begin{figure}[hbtp]
\includegraphics[scale=0.25]{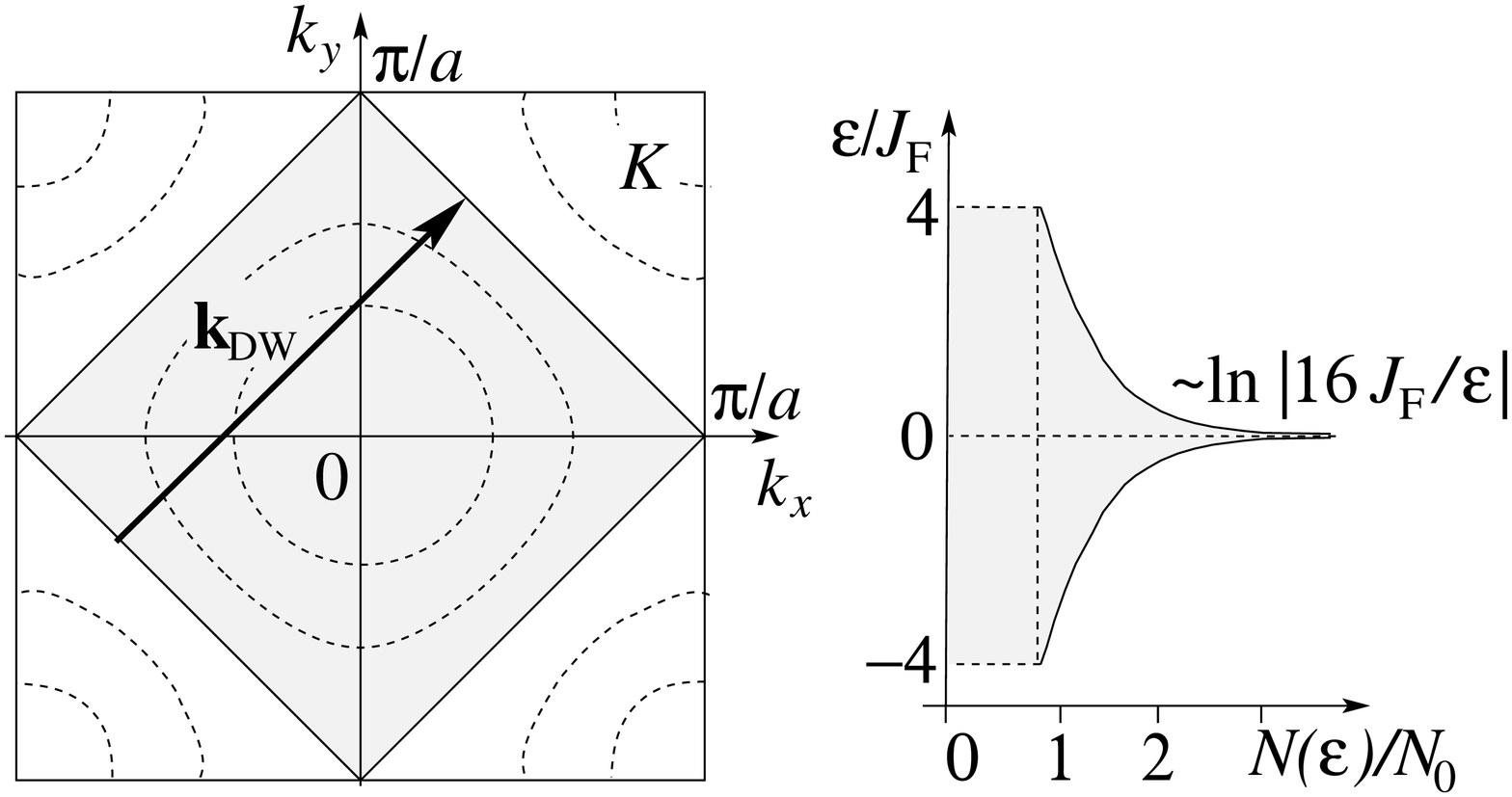}
  \caption{Left: first Brillouin zone $K$ and Fermi surface of
  2D fermions in an optical lattice. The solid lines denote the
  Fermi surface at half-filling. Right: density of states with
  logarithmic van Hove singularities $N(\epsilon) \sim
  N_{0} \ln | 16 J_{\rs F}/\epsilon|$.
  \label{fermisurface}}
\end{figure}

The Hamiltonian for interacting bosons and fermions subject to an
optical lattice takes the form $H = H_{\rm \scriptscriptstyle B} +
H_{\rm \scriptscriptstyle F}  + H_{\rm \scriptscriptstyle int}$
with ($\alpha= {\rm F,B}$)
\begin{eqnarray}
  H_{\rs \alpha}&=&\int d{\bf x} \:\:
     \psi^{+}_{\rm \scriptscriptstyle  \alpha}
     \left( - \frac{\hbar^{2}}{2m_{\rm \scriptscriptstyle \alpha}} \triangle +
      V_{\rm \scriptscriptstyle \alpha}({\bf x}) \right)
      \psi_{\rm \scriptscriptstyle  \alpha},  \label{startinghamilton}\\
  H_{\rm \scriptscriptstyle int}& = & \int d{\bf x} \left(g_{\rm
  \scriptscriptstyle FB}  \psi^{+}_{\rm \scriptscriptstyle B}
      \psi_{\rm \scriptscriptstyle  B} \psi^{+}_{\rm \scriptscriptstyle  F}
      \psi_{\rm \scriptscriptstyle  F} + \frac{1}{2}g_{\rm \scriptscriptstyle B}
      \psi^{+}_{\rm \scriptscriptstyle B}\psi^{+}_{\rm \scriptscriptstyle  B}
      \psi_{\rm \scriptscriptstyle  B}
      \psi_{\rm \scriptscriptstyle  B} \right). \nonumber
\end{eqnarray}
Here, we assume a repulsive interaction $g_{\rm \scriptscriptstyle
B}= 4 \pi a_{s} \hbar^{2}/m$ between the bosons, with the
scattering length $a_{s}>0$. The coupling $g_{\rs FB}=2 \pi a_{\rs
FB} \hbar^{2}/\mu$ accounts for the interaction between the
fermions and the bosons, with $\mu$ the relative mass and $a_{\rs
FB}$ the boson-fermion scattering length. Furthermore, we restrict
the analysis to spinless fermions; such a spinless fermionic atom
gas is naturally achieved via spin polarization. The $s$-wave
scattering length in the fermion system vanishes, while $p$-wave
scattering is small in general and is neglected in the following
analysis. The optical lattice with wave length $\lambda$ provides
an $a=\lambda/2$-periodic potential for the bosons and fermions
with $V_{\rm \scriptscriptstyle F,B}({\bf x}) =V_{\rm
\scriptscriptstyle F,B}[\sin^{2}( \pi x /a)+ \sin^{2}( \pi y /
a)]$. We expand the bosonic and fermionic field operators
$\psi_{\rm \scriptscriptstyle  B,F} $ in the Bloch wave functions
$v_{k,n}$ and $w_{k,n}$ of the single particle problem in a
periodic potential, (we restrict the analysis to the lowest Bloch
band)
\begin{equation}
    \psi_{\rm \scriptscriptstyle  B}({\bf x})=
    \!\!\sum_{{\bf k}\in K} b_{{\bf k}}
    w_{{\bf k}}({\bf x}), \hspace{5pt}
    \psi_{\rm \scriptscriptstyle  F}({\bf x})
    = \! \!\sum_{{\bf k}\in K} c_{{\bf k}} v_{{\bf k}}({\bf x}) .
\end{equation}
Here, $K$ denotes the first Brillouin zone, while $b_{{\bf k}}$
and $c_{{\bf k}}$ are the bosonic and fermionic annihilation
operators. For a strong optical lattice $V_{\rm \scriptscriptstyle
F,B}> E^{r}_{\rm \scriptscriptstyle F,B}=  2
\hbar^{2}\pi^{2}/\lambda^{2} m_{\rm \scriptscriptstyle F,B}$, the
restriction to the lowest Bloch band  is justified, and the
Hamiltonian simplifies to \cite{jaksch98}
\begin{eqnarray}
    H &= &\sum_{{\bf k} \in K}
    \epsilon_{\rs B}({\bf k}) b^{+}_{{\bf k}} b^{}_{{\bf k}}
    +\frac{U_{\scriptscriptstyle \rm B}}{2N}\!\! \sum_{\{{\bf k,k',q,q'}\}}
    b^{+}_{{\bf k}}b^{}_{{\bf k}'}  b^{+}_{{\bf q}} b^{}_{{\bf q}'} \nonumber\\
    & &\!\!\!\!+\sum_{{\bf q} \in K} \epsilon_{\scriptscriptstyle \rm
    F}({\bf q}) c^{+}_{{\bf q}}c^{}_{{\bf q}}
    + \frac{U_{\scriptscriptstyle \rm FB}}{N} \!\!\sum_{\{{\bf k,k',q,q'}\}}
    \!\!\!b^{+}_{ {\bf k}} b^{}_{{\bf k}'} c^{+}_{{\bf q}}c^{}_{{\bf
    q}'}, \label{tbhamilton}
\end{eqnarray}
with quantization volume $V=N a^{2}$ and $N$ the number of unit
cells (below, $n_{\rs F,B}$ denote the number of particles per
unit cell). The summation $\{ {\bf k,k',q,q'}\}$ is restricted to
${\bf k,k',q,q'} \in K$ with the momentum conservation ${\bf k}-
{\bf k}'+{\bf q}-{\bf q}'={\bf K}_{m}$; the reciprocal lattice
vector ${\bf K}_{m}$ accounts for Umklapp processes. Here,
$U_{\scriptscriptstyle \rm FB}= g_{\rs FB}\int d{\bf x}
|\widetilde{w}|^{2}|\widetilde{v}|^{2}$ and $U_{\scriptscriptstyle
\rm B}= g_{\rs B}\int d{\bf x}|\widetilde{w}|^{4}$, with
$\widetilde{w}({\bf x})$ and $\widetilde{v}({\bf x})$ the Wannier
functions associated with the Bloch band $w_{{\bf k}}$ and
$v_{{\bf k}} $ \cite{jaksch98}, while $\epsilon_{\rs F,B}({\bf
k})$ denote the energy dispersion of the fermions and bosons,
respectively. For a strong optical lattice only nearest neighbor
hopping survives and the dispersion relation takes the form
\begin{equation}
    \epsilon_{\scriptscriptstyle \rm F}({\bf q}) = -2 J_{\rm
    \scriptscriptstyle F} \left[ \cos \left(q_{x} a\right)
     +    \cos \left( q_{y} a\right) \right]
         \label{fermionicdispersion}
\end{equation}
with $J_{\rs F}$ the hopping energy. The Fermi surface at
half-filling $n_{\rs F}=1/2$ is shown in Fig.~\ref{fermisurface}
and exhibits perfect nesting for ${\bf k}_{\rs DW}=(\pi/a,\pi/a)$
and van Hove singularities at ${\bf k}=(0,\pm  \pi/a),(\pm
\pi/a,0)$. The bosonic energies are $ \epsilon_{\scriptscriptstyle
\rm B}({\bf q}) = 2 J_{\rm \scriptscriptstyle B} \left[2- \cos
\left(q_{x} a\right)- \cos \left( q_{y} a \right)\right]$ with
$\epsilon_{\rs B}(0)=0$.

Integrating out the fermions provides an effective interaction for
the bosons which depends on the temperature $T$ of the fermionic
atom gas. Within linear response theory, the boson density
operator $n_{\rs B}({\bf q})$ drives the fermionic density
$\langle n_{\rs F}({\bf q})\rangle= U_{\rm \scriptscriptstyle FB}
\chi(T,{\bf q}) n_{\rs B}({\bf q})$ with $\chi(T,{\bf q})$ the
fermionic response function. A perturbed fermionic density in turn
acts as a drive for the bosons, leading to an effective
boson-boson interaction
\begin{displaymath}
    H_{\rm \scriptscriptstyle int} =\!\frac{1}{2N} \!\! \sum_{\{{\bf
    k,k',q,q'}\}}\!\!\!\!\!
    \left[U_{\rm \scriptscriptstyle B}
    + U_{\rm \scriptscriptstyle FB}^{2} \: \chi(T,{\bf q-q'}) \right]
    b^{+}_{{\bf k}} b^{}_{{\bf k}'} b^{+}_{{\bf q}}b^{}_{{\bf q}'}
    \:.
    \label{interaction}
\end{displaymath}
The fermionic response is given by the Lindhard function
\begin{equation}
    \chi(T,{\bf q}) =  \int_{K}\frac{d{\bf k}}{v_{0}}
    \frac{f[\epsilon_{\rm \scriptscriptstyle F}({\bf k})]
    -f[\epsilon_{\rm \scriptscriptstyle F}({\bf k+q})] }
    {\epsilon_{\rm \scriptscriptstyle F}({\bf k})-
    \epsilon_{\rm \scriptscriptstyle F}({\bf k+q}) + i \eta}
    \label{lindhardfunction}
\end{equation}
with $v_{0}=( 2\pi/a)^{2}$ the volume of the first Brillouin zone.
The temperature $T$ enters via the Fermi distribution function
$f(\epsilon)=1/[1+\exp(\epsilon/T)]$ ($\mu_{\rs F}=0$ at half
filling); in our weak coupling analysis we are interested in
temperatures well below the superfluid transition temperature
$T_{\rs KT}$ of the bosons. Using the fermionic dispersion
relation (\ref{fermionicdispersion}), the Lindhard function
exhibits two logarithmic singularities. These singularities give
rise to instabilities in the system towards two new ground states:
the singularity at ${\bf q}=0$ induces an instability towards a
phase separated state, while the singularity at ${\bf k}_{\rs DW}$
induces an instability towards density wave formation and provides
a supersolid phase. In the following, we discuss these two
instabilities in detail.

For a fermionic system with a regular density of states, the
Lindhard function at ${\bf q}=0$  reduces to $\chi(T \rightarrow
0,0) =-N(0)$ at low temperatures, with $N(0)$ the fermionic
density of states. However, fermions on a square lattice within a
tight-binding scheme exhibit a logarithmic van Hove singularity in
the density of states, $N(\epsilon) = N_{0}
K[\sqrt{1-\epsilon^{2}/16J_{\rs F}^{2}}] \sim N_{0} \ln|16 J_{\rs
F}/\epsilon | $ with $N_{0}= 1/(2\pi^{2} J_{\rs F})$ and $K[k]$
the complete elliptic integral of the first kind
\cite{gradshteyn}, see Fig.~\ref{fermisurface}. The response at
half-filling $n_{\rs F}=1/2$ then behaves as
\begin{equation}
    \chi(T\rightarrow 0,0)= \int d\epsilon N(\epsilon)
     \partial_{\epsilon} f(\epsilon)
    \sim-N_{0}
    \ln\frac{16 c_{1} J_{\rm \scriptscriptstyle F}}{T} \label{q0singularity}
\end{equation}
with $c_{1} =2\exp(C)/\pi\approx 1.13$.

The coupling between the bosons and the fermions induces an
attraction between the bosons, and the effective long distance
scattering parameter takes the form
$U_{\rm \scriptscriptstyle eff}=  U_{\rm \scriptscriptstyle
B} + U_{\rm \scriptscriptstyle FB}^{2} \: \chi(T,0)$.
%
%
The thermodynamic stability of a superfluid condensate at low
temperatures requires a positive effective interaction $U_{\rm
\scriptscriptstyle eff}>0$ and the condition $U_{\rm
\scriptscriptstyle eff}(T_{\rs PS})=0$ defines the critical
temperature $T_{\rs PS}$ for phase separation. Using
Eqs.~(\ref{q0singularity}),
we find for the critical temperature
\begin{equation}
    T_{\rm \scriptscriptstyle PS} = 16 c_{1}
        J_{\rm \scriptscriptstyle F} \exp\left[-
        1/\lambda_{\rs FB} \right]\label{TPS}
\end{equation}
with $\lambda_{\rs FB} = U_{\rs FB}^{2}N_{0}/U_{\rs B}\ll 1 $ the
ratio between  the induced attraction and the intrinsic repulsion
between the bosons. Below the critical temperature $T_{\rs PS}$,
the effective interaction $U_{\rs eff}$ turns negative, providing
a negative compressibility and rendering the Bose system unstable.
The new ground state with fixed averaged densities $n_{\rs B}$ and
$n_{\rs F}$ exhibits phase separation with areas of increased and
decreased local densities coexisting, cf.\ Fig.\ 1.

Note, that this transition towards phase separation exhibits two
major differences as compared to the phase separation discussed in
Refs.~\onlinecite{molmer98,viverit00}. First, our phase separation
appears as an instability, i.e., for arbitrary small coupling
$U_{\rs FB}$ between the bosons and fermions. Second, the
increase/decrease in the bosonic density drives the fermionic
density away from half-filling, providing a regular $\chi(T,0)$
which stabilizes the system.

This phase separation is in competition with a second instability
in the system triggered by the singularity in the Lindhard
function at ${\bf k}_{\rs DW}$. Using (\ref{lindhardfunction}) and
the symmetry $\epsilon_{\scriptscriptstyle \rm F}({\bf q}+{\bf
k}_{\rs DW})\!= - \epsilon_{\scriptscriptstyle \rm F}({\bf q})$,
the Lindhard function becomes
\[
    \chi(T,{\bf k}_{\rs DW})\! = \!\!\int\!\! d\epsilon N(\epsilon)
     \frac{\tanh\left({\epsilon}/{2T}\right)}
     {-2\:\epsilon}\!\sim\!-\frac{N_{0}}{2}
    \!\left[\ln  \frac{16 c_{1}
    J_{\rm \scriptscriptstyle F}}{T}\right]^{2}\!\!.
\]
The combination of van Hove singularities and perfect nesting
produces the $[\ln T]^{2}$ singular behavior known to produce a
$T_c$-enhancement in superconductivity \cite{hirsch86}. Within
Bogoliubov theory, the bosonic quasi-particle spectrum becomes
\begin{equation}
    E_{\rm \scriptscriptstyle B}({\bf q}) =
    \sqrt{\epsilon_{\rm \scriptscriptstyle B}^{2}({\bf q}) + 2 n_{\rs B}
    \epsilon_{\rm \scriptscriptstyle B}({\bf q})
    \left[U_{\rm \scriptscriptstyle B}
    + U_{\rm \scriptscriptstyle FB}^{2} \: \chi(T,{\bf q})
    \right]}.
\end{equation}
The induced attraction between the bosons reduces the energy of
quasi-particles at ${\bf k}_{\rs DW}$ which vanishes ($E_{\rs
B}({\bf k}_{\rs DW}) =0$) at the critical temperature
\begin{equation}
    T_{\rm \scriptscriptstyle DW} = 16 c_{1} \: J_{\rm \scriptscriptstyle F}
    \exp\left[ - \sqrt{(2+t_{\rs B})/\lambda_{\rs FB}}\right]
    \label{CDWtemp}
\end{equation}
with $t_{\rs B}= 8J_{\rs B}/n_{\rs B}U_{\rs B}$ the ratio between
the kinetic and the interaction energy of the bosons (weak
coupling requires $t_{\rs B}\gg \lambda_{\rs FB}$). Below this
critical temperature, the boson mode $b_{{\bf k}_{\rs DW}}$
becomes macroscopically occupied and its interference with the
condensate produces a bosonic density wave, see below. In this new
phase, a (quasi-)condensate characterized by an off-diagonal
(quasi-)long-range order with a finite superfluid stiffness
coexists with a density wave providing diagonal long-range order,
thus establishing a supersolid phase. Note, that decreasing the
hopping $t_{\rs B}$ drives a superfluid to Mott-insulator
transition for commensurate densities \cite{jaksch98} below
$t_{\rs B}<t_{\rs SF-MI}\approx 1/3$, providing an additional
competing phase at strong coupling $t_{\rs B}<1$.

Next, we study the supersolid phase within a mean field
description. We introduce the mean fields $\langle b_{0} \rangle=
\sqrt{n_{0}N} \exp(i \varphi_{0})$ and $\langle b_{{\bf k}_{\rs
DW}}\rangle = (\Delta /2 U_{\rm \scriptscriptstyle FB})
\sqrt{N/n_{0}}\exp(i \varphi)$ and neglect thermal excitations of
bosonic quasi-particles at $ T \ll T_{\rs KT}$, implying the
constraint $n_{\rs B}= n_{0} + \Delta^{2}/(4n_{0} U^{2}_{\rs
FB})$. The bosonic density takes the form (to be evaluated at
lattice sites)
\begin{equation}
        n_{\rm \scriptscriptstyle B}(x,y)=n_{\rm \scriptscriptstyle
        B}+ \frac{ \Delta \cos \theta}{ U_{\rm \scriptscriptstyle FB}}
        \left[\cos \frac{\pi x}{a} \cos \frac{\pi y}{a}\right]
        \label{densitywave}
\end{equation}
with $\theta= \varphi_{0} - \varphi$; this bosonic density wave
appears as a result of the interference between the two
condensates $\langle b_{0} \rangle$ and $\langle b_{{\bf k}_{\rs
DW}}\rangle$. Neglecting terms independent on $\Delta$, the
Hamiltonian (\ref{tbhamilton}) per unit cell reduces to
\begin{equation}
    \frac{H}{N}= 2 J_{\rs B} \frac{\Delta^{2}}
    { n_{\rm \scriptscriptstyle B}U_{\rs FB}^{2}}
     +\frac{ U_{\rm \scriptscriptstyle B}
     \Delta^{2} \cos^{2} \theta}{2 U_{\rm \scriptscriptstyle FB}^{2}}
     + \frac{H_{\rs F}}{N}
     + o(\Delta^{4}).
\end{equation}
The first and second terms describe  the increase in the kinetic
and interaction energies of the bosons, while  $H_{\rs F}$
accounts for the nesting of fermions with ${\bf q} \in K$ and
${\bf q}'= {\bf q} - {\bf k}_{\rs DW}+{\bf K}_{m}$ (the reciprocal
lattice vector ${\bf K}_{m}$ ensures the constraint ${\bf q}'\in
K$)
\begin{equation}
    H_{\rs F} =\frac{1}{2}\sum_{{\bf q} \in K}\!\!
    \left(\! \begin{array}{cc} c_{{\bf q}}^{+}\:,&\!\!
    c_{{\bf q}'}^{+}
    \end{array}\!\right)\!
    \left( \!\begin{array}{cc} \epsilon_{\rm \scriptscriptstyle F}({\bf q}) &  \Delta \cos\theta \\
    \Delta \cos\theta &  \epsilon_{\rm \scriptscriptstyle F}({\bf
    q}')
    \end{array}\!\right)\!
       \left( \!\begin{array}{c} c^{}_{{\bf q}}\\
    c^{}_{{\bf q}' }
    \end{array}\!\right).
\end{equation}
Diagonalizing, we obtain the fermionic quasi-particle excitation
spectrum $\widetilde{\epsilon}_{\rm \scriptscriptstyle F}({\bf
k},\Delta)= \pm [\epsilon_{\rm \scriptscriptstyle F}^{2}({\bf
k})+\cos^{2}\theta\Delta^{2}]^{1/2}$. Minimizing the thermodynamic
potential $\Omega(T,\Delta, \theta)$ provides the constraint
$\theta = s \pi$, with $s$ an integer, and the self-consistency
relation ($\partial_{\Delta} \Omega=0$)
\begin{equation}
    \frac{1}{\lambda_{\rs FB}}\left(2+t_{\rs B}\right) =
    \frac{1}{N_{0}}\int_{K}\frac{d{\bf k}}{v_{0}} \frac{\tanh
\left[\widetilde{\epsilon}_{\rm \scriptscriptstyle F}({\bf
k},\Delta)/2 T\right]}{\widetilde{\epsilon}_{\rm
\scriptscriptstyle F}({\bf k},\Delta)} \: .
\end{equation}
Setting $\Delta\!\! =\!\! 0$, we reproduce the critical
{temperature} (\ref{CDWtemp}). Using the density of states
$N_{\Delta}(\epsilon)=N(\sqrt{\epsilon^2-\Delta^2})$
$|\epsilon|/\sqrt{\epsilon^2+\Delta^{2}}$, the gap at $T=0$
becomes
\begin{equation}
    \Delta(0)= 32 J_{\rm \scriptscriptstyle F}
    \exp\left[ -\sqrt{(2+t_{\rs B})/\lambda_{\rs FB}} \right]
\end{equation}
and we obtain the standard BCS relation $2 \Delta(0)/ T_{\rm
\scriptscriptstyle DW}= 2 \pi/e^C\approx 3.58$. The mean field
$\langle b_{0}\rangle= \sqrt{n_{0}N} \exp[i \varphi_{0}]$ breaks
the continuous $U(1)$ symmetry of the system and describes the
off-diagonal quasi-long-range order. The excitation spectrum
exhibits a linear dispersion around $q=0$ and the quasi-long-range
order is sufficient to provide a finite superfluid stiffness
\cite{leggett70}. On the other hand, the mean field $\langle
b_{{\bf k}_{\rs DW}}\rangle$ characterizes a commensurate density
wave. Its phase is locked to the condensate phase via the
constraint $\theta = s \pi$ and the excitation spectrum around
${\bf q}= {\bf k}_{\rs DW}$ is gapped. The transition breaks the
discrete translation symmetry and establishes diagonal long-range
order; breaking a discrete symmetry, diagonal long-range order can
exist even at finite temperatures.

The competition between the two instabilities at ${\bf q}=0$ and
${\bf k}_{\rs DW}$ provides the phase diagram as shown in
Fig.~\ref{phasedia}. For $T_{\rs DW}>T_{\rs PS}$ the system
undergoes a transition into a supersolid phase at $T_{\rs DW}$;
the gap in the fermionic excitation spectrum then removes the
instability towards phase separation luring at lower temperatures.
In turn, for $T_{\rs PS}>T_{\rs DW}$ the instability towards phase
separation wins over the density wave formation and drives the
fermionic density away from half-filling; the nesting at ${\bf
k}_{\rs DW}$ is quenched and the instability towards density wave
formation disappears. The projection of the critical line $T_{\rs
PS} = T_{\rs DW}$ onto the $\lambda_{\rs FB}$-$1/t_{\rs B}$-plane
satisfies the relation $1/t_{\rs B}=\lambda_{\rs
FB}/(1-2\lambda_{\rs FB})$; with decreasing coupling $\lambda_{\rs
FB}$ we enter the supersolid phase, while increasing the hopping
$t_{\rs B}$ drives the system towards phase separation.

\begin{figure}[hbtp]
\vspace{-0.cm}
\includegraphics[scale=0.58]{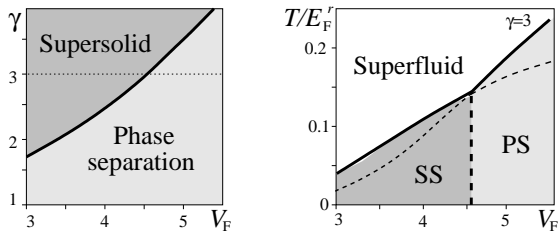}
  \caption{ \label{realphasedia} Left:
  $V_{\rs F}$-$\gamma$ phase diagram at low
  temperatures. Increasing the strength of the optical lattice in
  an experiment allows to drive the transition from a supersolid
  to a phase separated state. Right: Transition temperatures
  for $\gamma=3$ ($E_{\rs F}^{r}=347~{\rm nK}$).}
\end{figure}

Finally, we estimate the relevant experimental parameters for an
atomic mixture of fermionic $^{40}$K and bosonic $^{87}$Rb with
scattering lengths $a_{\rs B}= 5.77~{\rm nm}$ and $a_{\rs FB}
\approx 15~{\rm nm}$ \cite{roati02}. The 2D setup is realized
through application of an anisotropic 3D optical lattice ($\lambda
= 830~{\rm nm}$ and $V_{\rs F}/V_{\rs B}\approx 3/7$), with
$V_{\rs F}^{z} \gg V_{\rs F}$ and $V_{\rs B}^{z} \gg V_{\rs B}$
quenching the inter-plane hopping (we express $V_{\rs F, B}$ and
$V_{\rs F, B}^{z}$ via the recoil energies $E_{\rs F,
B}^{r}=2\pi^2 \hbar^2/\lambda^2m_{\rs F,B}$). The hopping
amplitudes $J_{\rs F,B}$ derive from the 1D Mathieu equation,
\begin{equation}
    {J_{\rs F, B}} = ({4}/{\sqrt{\pi}})
     E_{\rs F, B}^{r} V_{\rs F, B}^{3/4} \exp\bigl(- 2 \sqrt{V_{\rs
    F, B}}\bigr),
\end{equation}
while the interactions $U_{\rs FB}$ and $U_{\rs B}$ are given as
\cite{jaksch98}
\begin{equation}
    \frac{U_{\rs FB}}{E_{\rs F}^{r}}
    = 8 \sqrt{\pi}\frac{1+m_{\rs F}/m_{\rs
    B}}{(1+\sqrt{V_{\rs F}/V_{\rs B}})^{3/2}}
    \frac{a_{\rs FB}}{\lambda\gamma}
    \left(V_{\rs F}^{z}\right)^{1/4} V_{\rs F}^{1/2}
 \label{fermiboseinteraction}
\end{equation}
and $U_{\rs B}/E_{\rs B}^{r} = 4 \sqrt{2\pi}(a_{\rs B}/\lambda
\gamma) \left(V_{\rs B}^{z}\right)^{1/4} V_{\rs B}^{1/2}$. Using a
finite angle between the laser beams producing the standing light
waves, we allow to change the relative size of the in- and
out-of-plane lattice constants $a$ and $a_{z}$. The parameter
$\gamma=2 a_{z}/\lambda$ then denotes the increase in the unit
cell volume and allows to tune the interaction strengths $U_{\rs
B}$ and $U_{\rs FB}$ independent on $J_{\rs F,B}$ (alternatively,
Feshbach resonances allow to tune $a_{\rs FB}$). Fixing $V_{\rs
F}^{z} = 20$, $n_{\rs F}=1/2$, and $n_{\rs B}=3/2$, we obtain for
$\gamma=3$ and $V_{\rs F}=4.5$ the coupling parameters
$\lambda_{\rs FB}\approx 0.39$, $t_{\rs B}\approx 0.55$ ($T_{\rs
DW}\approx 48~{\rm nK}$); we enter a regime at the border of
validity of our weak coupling analysis. Using the above estimates,
the $V_{\rs F}$-$\gamma$ phase diagram is shown in
Fig.~\ref{realphasedia}; we find that changing the strength of the
optical lattice $V_{\rs F}$ allows to drive the transition from
the supersolid to a phase separated state.
 The supersolid state is easily
detected via the usual coherence peak of a bosonic condensate in
an optical lattice and the additional appearance of coherence
peaks at ${\bf k}_{\rs DW}$; using the above parameters, the
weight of these additional peaks involve $15\%$ of the total
particle number.

In conclusion, we have identified a new supersolid phase in a 2D
Fermi-Bose gas mixture subject to an optical lattice; the bosons
then play the role of the phonons in a condensed matter system.
The perfect Fermi-surface nesting leads to the appearance of a
fermionic density wave and the condensation of the bosons at ${\bf
k}_{\rs DW}$. The interference of this ${\bf k}_{\rs
DW}$-condensate with the usual ${\bf q}=0$-condensate establishes
the bosonic density wave characteristic of the supersolid phase.

{\acknowledgments We thank T.~M.\ Rice, T.\ Esslinger, M.\ K\"ohl,
and M.\ Troyer for stimulating discussions.}


\end{document}